\documentclass[aps,amsfonts,twocolumn,pre]{revtex4}
\usepackage{epsfig,graphicx,amsmath,amsfonts,eucal,bm,mathrsfs}
\newcommand{\B}[1]{{\bm{#1}}}
\newcommand{\C}[1]{{\mathcal{#1}}}
\usepackage[latin1]{inputenc}
\newcommand{\beq}{\begin{equation}}
\newcommand{\eeq}{\end{equation}}
\newcommand{\bea}{\begin{eqnarray}}
\newcommand{\eea}{\end{eqnarray}}

\begin{document}
\title{Elasticity with Arbitrarily Shaped Inhomogeneity}
\author{Joachim Mathiesen$^{(1)}$, Itamar Procaccia$^{(2)}$ and Ido Regev$^{(2)}$}
\affiliation{$^{(1)}$ Physics of Geological Processes, Univ. of Oslo, Postbox 1048 Blindern, N-0316 Oslo, Norway\\$^{(2)}$ Dept. of Chemical Physics, The Weizmann Institute of Science, Rehovot 76100, Israel}

\begin{abstract}
A classical problem in elasticity theory involves an inhomogeneity embedded in a material
of given stress and shear moduli. The inhomogeneity is a region of arbitrary shape
whose stress and shear moduli differ from those of the surrounding medium.  In this paper we present a new, semi-analytic method for finding the stress tensor for an infinite plate with such an inhomogeneity.
The solution involves two conformal maps, one from the inside and the second from the outside of the unit circle to the inside, and respectively outside, of the inhomogeneity. The method provides
a solution by matching the conformal maps on the boundary
between the inhomogeneity and the surrounding material. This matching converges well  only for relatively mild distortions of  the unit circle due to reasons which will be discussed in the article. We provide a comparison of the present result to known previous results.
\end{abstract}
\maketitle

\section{Introduction}

Elasticity theory in homogeneous materials is a well developed subject. Much less
is known about inhomogeneous materials where the solution of the basic equations of
elasticity becomes very involved. In this paper we focus on a material which consists
of one finite area (of arbitrary shape) in which a material with given elastic properties
is embedded in an infinite sheet of material of different elastic properties. This situation
is known as ``elastic inhomogeneity" and it appears in a variety of solid mechanical contexts. Previous studies have concentrated on solving this problem for the relatively symmetric case of an ellipse \cite{54Har} where it was solved analytically. In other cases the problem was solved for small perturbations to the circle \cite{90Gao}.
\begin{figure}
\centering \epsfig{width=.45\textwidth,file=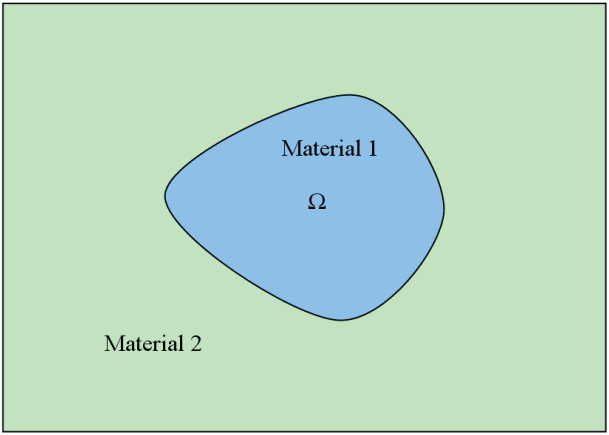}
 \caption{The
region $\Omega$}\label{patch}
\end{figure}
Mathematically the problem is set as follows, and see Fig. \ref{patch}.
A patch of material of type 1 occupies an area $\Omega$ and is delineated by a sharp boundary  which
will be denoted $\partial\Omega$. The rest of the infinite plane is made of material of type 2.
The material is subjected to forces at infinity (and see below the precise boundary conditions), and
is therefore deformed. Before the deformation each point of the material
is assigned a point $\B r$ in the two-dimensional plane. The forces at infinity result in
a displacement of the material points to a new equilibrium position $\B r'$. The displacement
field $\B u(\B r)$ is defined as
\begin{equation}
\B u(\B r) \equiv \B r'-\B r \ .
\end{equation}
The strain field is defined accordingly as
\begin{equation}
\epsilon_{ij} \equiv  [\partial_i \B u_j+\partial_j \B u_i]/2 \ .
\end{equation}
In the context  of linear elasticity in isotropic materials one then introduces the
stress field according to Hook's law
\begin{equation}
\sigma_{ij}=2\mu_i\epsilon_{ij}+\lambda_i\delta_{ij}\epsilon_{kk}
\end{equation}
Where $\lambda_i=\mu_i/(\frac{1}{2\nu_i}-1)$. $\mu_i$ and $\nu_i$
take on different values $\mu_1$, $\nu_1$ in $\Omega$ and $\mu_2$,
$\nu_2$ in the rest of the material. In equilibrium the stress
tensor should be divergenceless
$\frac{\partial\sigma_{ij}}{\partial x_j}=0$ at each point in the
sheet. By defining the  stress (or Airy) potential U:
\begin{equation}
\sigma_{xx}=\frac{\partial^2 U}{\partial y^2} \ ; \sigma_{xy}=-
\frac{\partial^2 U}{\partial x\partial y} \ ;
\sigma_{yy}=\frac{\partial^2 U}{\partial x^2} \ , \label{sigU}
\end{equation}
the former equation for the stress tensor becomes a partial
differential equation for the stress potential:
\begin{equation}
\nabla^2 \nabla^2 U(x,y) =0 \ . \label{bilaplace}
\end{equation}
This equation, which is known as the bi-Laplace or the bi-harmonic equation is conveniently
solved as a non-analytic combination of analytic funcitons. To this aim we introduce the complex
notation $z\equiv x+iy$, and note the general solutions of Eq. (\ref{bilaplace}) in the form
\begin{equation}
U(x,y)= \Re [\bar z \tilde\varphi(z)+\eta(z)] \ ,
\label{Uphichi}
\end{equation}
where $\bar z\equiv x-iy$ and  $\tilde\varphi(z)$ and $\eta(z)$ are any two holomorphic functions.
What remains to do in any particular problem is to find the unique analytic functions such that
the stress tensor satisfies the boundary conditions. This stress tensor is
determined by the two holomorphic functions as:
\begin{eqnarray}
\sigma_{yy}(x,y)&=&\Re [2 \tilde\varphi'(z)+ \bar z\tilde\varphi''(z)+\eta''(z)]\nonumber\\
\sigma_{xx}(x,y)&=&\Re [2 \tilde\varphi'(z)-\bar z\tilde\varphi''(z)-\eta''(z)]\nonumber\\
\sigma_{xy}(x,y)&=&\Im [\bar z\tilde\varphi''(z)+\eta''(z)].
\end{eqnarray}
We define for convenience:
\begin{equation}
\tilde{\psi}(z)=\eta'(z)
\end{equation}
And then:
\begin{eqnarray}
\sigma_{yy}(x,y)&=&\Re [2 \tilde\varphi'(z)+ \bar z\tilde\varphi''(z)+\tilde\psi'(z)]\nonumber\\
\sigma_{xx}(x,y)&=&\Re [2 \tilde\varphi'(z)-\bar z\tilde\varphi''(z)-\tilde\psi'(z)]\nonumber\\
\sigma_{xy}(x,y)&=&\Im [\bar z\tilde\varphi''(z)+\tilde\psi'(z)].
\end{eqnarray}

Note that the stress tensor is determined by derivatives of the holomorphic
functions, and not by the functions themselves. This leaves
us with some freedom, since the functions can be changed with the following gauge:
\begin{eqnarray}
\tilde\varphi_i &\rightarrow& \tilde\varphi_i +iC_iz+\alpha_i+i \beta_i\nonumber\\
\tilde\psi_i &\rightarrow& \tilde\psi_i +\gamma_i+i \delta_i \
, \quad\tilde\psi\equiv \eta'
\end{eqnarray}
As we shall see below, not all these gauge freedoms are
true freedoms once we introduce the boundary and continuity conditions. Since the
elastic properties are different inside and outside $\Omega$, the
potential functions will be different in the two regions:
$\tilde\varphi_1$ and $\tilde\psi_1$ which are defined on $\Omega$
and $\tilde\varphi_2$ and $\tilde\psi_2$ which are defined on
$\mathbb{C}\backslash\Omega$. Nevertheless we will demand continuity of
the physical fields. In particular the normal force
\begin{equation}
\B {\sigma}\cdot
\textbf{n}=\sigma_{xn}+i \sigma_{yn}
\end{equation}
and the displacement $\textbf{u}(\textbf{r})$ must be continues across the interface (by Newton's third law) in
the absence of surface tension. Therefore the continuity
conditions are:
\begin{equation}
\sigma^{(1)}_{xn}+i\sigma^{(1)}_{yn}=\sigma^{(2)}_{xn}+i\sigma^{(2)}_{yn} \ , \label{constress}
\end{equation}
\begin{equation}
u^{(1)}_x+i u^{(1)}_y=u^{(2)}_x+i u^{(2)}_y \ .
\label{condis}
\end{equation}
The continuity conditions for the stress, can be rewritten as:
\begin{equation}
\frac{d}{ds}\left(\frac{\partial U_1}{\partial x} +i\frac{\partial
U_1}{\partial y}\right)=\frac{d}{ds}\left(\frac{\partial
U_2}{\partial x} +i\frac{\partial U_2}{\partial y}\right)
\end{equation}
or, after integrating:
\begin{equation}
\frac{\partial U_1}{\partial x} +i\frac{\partial U_1}{\partial
y}=\frac{\partial U_2}{\partial x} +i\frac{\partial U_2}{\partial
y}+\C C\ ,
\end{equation}
where $\C C$ is a complex constant of integration. In terms of the holomorphic functions, the condition (\ref{constress}) translates to \cite{53Mus}:
\begin{widetext}
\begin{equation}
\tilde\varphi^{(1)}(z)+z\overline{\tilde\varphi'^{(1)}(z)}+\overline{\tilde\psi^{(1)}(z)}=\tilde\varphi^{(2)}(z)+z\overline{\tilde\varphi'^{(2)}(z)}+\overline{\tilde\psi^{(2)}(z)}+\C C \ ,
\end{equation}
and the condition (\ref{condis}) becomes:
\begin{equation}
\frac{[\kappa_1\tilde\varphi^{(1)}(z)\!-\!z\overline{\tilde\varphi'^{(1)}(z)}\!-\!\overline{\tilde\psi^{(1)}(z)}]}{\mu_1}=\frac{[\kappa_2\tilde\varphi^{(2)}(z)\!-\!z\overline{\tilde\varphi'^{(2)}(z)}\!-\!\overline{\tilde\psi^{(2)}(z)}]}{\mu_2} \ .
\end{equation}
where $\kappa_i=(3-\nu_i)/(1+\nu_i)$.
\end{widetext}
In addition to these continuity conditions on $\partial \Omega$ we need to specify boundary conditions
at infinity. We choose
\begin{equation}
\sigma_{xx}(\infty)=0 \ , \quad
\sigma_{yy}(\infty)=\sigma_\infty  \ , \quad
 \sigma_{xy}(\infty)=0 \ . \label{bcinfty}
\end{equation}

\section{Solution in terms of conformal maps}

Solutions to the problem of finding the stress field {\em outside} a given domain
using conformal maps were described for example in \cite{Barra}. Here we need to 
solve for the stress field both inside and outside the given domain.
In the following we assume that the center of coordinates is inside
$\Omega$ and the point at infinity is outside $\Omega$. Since the
stress functions are holomorphic in their domains of definition, we
can expand them in the appropriate laurent series, which for the
functions with superscript (1) is of the form
\begin{eqnarray}
\tilde\varphi^{(1)}(z) &=&\tilde\varphi^{(1)}_0
+\tilde\varphi^{(1)}_{1}z+\tilde\varphi^{(1)}_{2}z^2+\cdots \ , \nonumber\\
\tilde\psi^{(1)}(z) &=&\tilde\psi^{(1)}_0
+\tilde\psi^{(1)}_{1}z+\tilde\psi^{(1)}_{2}z^2+\cdots \ .
\label{Laurent1}
\end{eqnarray}
i.e. we have no poles at the origin. For the outside domain (functions
 with superscript (2)) the most general expansions in agreement with the
 boundary conditions (\ref{bcinfty}) are of the form

\begin{eqnarray}
\tilde\varphi^{(2)}(z) &=&\tilde\varphi^{(2)}_1 z + \tilde\varphi^{(2)}_0
+\tilde\varphi^{(2)}_{-1}/z+\tilde\varphi^{(2)}_{-2}/z^2+\cdots \ , \nonumber\\
\tilde\psi^{(2)}(z) &=&\tilde\psi^{(2)}_1 z + \tilde\psi^{(2)}_0
+\tilde\psi^{(2)}_{-1}/z+\tilde\psi^{(2)}_{-2}/z^2+\cdots \ .
\label{Laurent2}
\end{eqnarray}
i.e we have a pole of order 1 at infinity. Accordingly, the
leading terms of Eqs. (\ref{Laurent2}) are determined by the boundary
conditions.  We now use one of the gauge freedoms to eliminate the imaginary part of $\varphi^{(2)}$ and write:
\begin{equation}
 \tilde\varphi^{(2)}_1=\frac{\sigma_{\infty}}{4}\ ;\quad
 \tilde\psi^{(2)}_1=\frac{\sigma_{\infty}}{2} \ .
\end{equation}

The standard way to proceed \cite{53Mus} would be to substitute
the series expansions in the continuity conditions and find the
linear equations that determine all the coefficients by equating
terms of the same order in $z$. However this cannot be done in general
since the functions $z^n$ are not orthogonal on arbitrary contours
$\partial \Omega$. To overcome this, one maps the
regions $\Omega$ and $\mathbb{C}\backslash\Omega$ into the
interior and exterior of the unit circle, respectively. That is, we need two holomorphic, invertible (and thus conformal) functions, one is
\begin{equation}
z=\Phi(\omega)
\end{equation}
which maps the exterior of the unit circle into $\mathbb{C}\backslash\Omega$,
and the other is
\begin{equation}
z=\Lambda(\zeta)
\end{equation}
which maps the unit disk into $\Omega$.  Since they are both
invertible they have inverse functions which we denote
\begin{equation}
\zeta=\Lambda^{-1}(z)
\end{equation}
and
\begin{equation}
\omega=\Phi^{-1}(z) \ .
\end{equation}

Now we express the functions  $\tilde \varphi^{(i)}$ and $\tilde \psi^{(i)}$ in terms of $\omega$ and
$\zeta$ and then expand them on the boundary of the unit circle. This expansion
will be a Fourier series where the powers of $\omega$ or $\zeta$ satisfies the orthogonality relation:
\begin{equation}
\frac{1}{2\pi}\int_0^{2\pi}
e^{n i\theta}e^{-m i\theta}=\delta_{n,m} \ . \label{ortho}
\end{equation}
We have here used $e^{i \theta}$ to represent either $\omega$ or $\zeta$. The orthogonality allows us to equate the coefficients of the series term by term. We define:
\begin{equation}
\tilde\varphi^{(2)}(z)\equiv  \varphi^{(2)}\left(\Phi^{-1}(z)\right) \ , \quad
\tilde\psi(z)^{(2)}\equiv  \psi^{(2)}\left(\Phi^{-1}(z)\right) \ . \label{ppp}
\end{equation}
and:
\begin{equation}
\tilde\varphi^{(1)}(z)\equiv \varphi^{(1)}\left(\Lambda^{-1}(z)\right) \ ,
\quad \tilde\psi^{(1)}(z)\equiv \psi^{(1)}\left(\Lambda^{-1}(z)\right) \ .
\label{ppp}
\end{equation}
We can expand $\varphi_i$ and $\psi_i$ in terms of $\omega$
and $\zeta$ on the unit circle. Since the original functions were holomorphic and
meromorphic in the original domains, the functions:
\begin{equation}
\varphi^{(2)}(\omega)\equiv  \tilde\varphi^{(2)}\left(\Phi(\omega)\right)
\ , \quad \psi^{(2)}(\omega)\equiv
\tilde\psi^{(2)}\left(\Phi(\omega)\right) \ . \label{ppp}
\end{equation}
and:
\begin{equation}
\varphi^{(1)}(\zeta)\equiv \tilde\varphi^{(1)}\left(\Lambda(\zeta)\right)
\ , \quad \psi^{(1)}(\zeta)\equiv
\tilde\psi^{(1)}\left(\Lambda(\zeta)\right) \ . \label{ppp}
\end{equation}
are holomorphic inside and outside the unit disc, respectively.
Therefore we can expand in terms of $\omega$ and
$\zeta$:
\begin{eqnarray}
\varphi^{(1)}(\zeta) &=&\varphi^{(1)}_0
+\varphi^{(1)}_{1}\zeta+\varphi^{(1)}_{2}\zeta^2+\cdots \ , \nonumber\\
\psi^{(1)}(\zeta) &=&\psi^{(1)}_0
+\psi^{(1)}_{1}\zeta+\psi^{(1)}_{2}\zeta^2+\cdots \ . \label{Laurentp1}
\end{eqnarray}
\begin{eqnarray}
\varphi^{(2)}(\omega) &=&\varphi^{(2)}_1 \omega + \varphi^{(2)}_0
+\varphi^{(2)}_{-1}\omega^{-1}+\varphi^{(2)}_{-2}\omega^{-2}+\cdots \ , \nonumber\\
\psi^{(2)}(\omega) &=&\psi^{(2)}_1 \omega + \psi^{(2)}_0
+\psi^{(2)}_{-1}\omega^{-1}+\psi^{(2)}_{-2}\omega^{-2}+\cdots \ .
\label{Laurentp2}
\end{eqnarray}
We now assume that the map of the exterior domain, $\Phi$,  maps the
point at infinity to infinity. That is, it will have a Laurent series
on the form
\begin{equation}
\Phi(\omega ) = F_1\omega + F_0
+F_{-1}\omega^{-1}+F_{-2} \omega^{-2}+\cdots \ . \label{Laurent}
\end{equation}
From this we get the following relations (after substituting and
taking the limit $\omega\rightarrow\infty$):
\begin{eqnarray}
\varphi^{(2)}_1&=&F_1\frac{\sigma_{\infty}}{4}\ ;\quad
\psi^{(2)}_1=F_1\frac{\sigma_{\infty}}{2} \quad
\end{eqnarray}
we can also use the last two freedoms to choose
$\tilde\varphi^{(2)}_0= -F_0\tilde\varphi^{(2)}_1$ such that
$\varphi^{(2)}_0=0$. In the interior domain, the functions  $\tilde
\varphi^{(1)}$ and $\tilde \psi^{(1)}$ also have five freedom. However, the
requirement of continuity of the displacement field $\B u$ across
the boundary $\partial\Omega$ removes three of these freedoms.
This continuity was expressed by Eq. (\ref{condis}). Applying the
apparent gauge freedoms on the LHS of that equation and then
subtracting the resulting equation from Eq. (\ref{condis}) we find
the three conditions
\begin{equation}
C_1=0\ , \quad \kappa_1\alpha_1=\gamma_1\ , \quad \kappa_1\beta_1=-\delta_1 \ .
\end{equation}

Using the remaining two freedoms, we can eliminate the constant term in the expansion of $\psi^{(1)}$ by setting $\psi^{(1)}_0=0$. Note that this is possible only when we choose $\Lambda(\zeta)$ such that $\Lambda(0)=0$. We may always define our mapping $\Lambda$ such that this is satisfied. In terms of the conformal
maps we transform the boundary conditions into
\begin{eqnarray}
&&\varphi^{(1)}(\zeta)+\frac{\Lambda(\zeta)}{\overline{\Lambda'(\zeta)}}{\overline{\varphi'^{(1)}(\zeta)}}+\overline{\psi^{(1)}(\zeta)}=\nonumber\\&&\varphi^{(2)}(\omega)+\frac{\Phi(\omega)}{\overline{\Phi'(\omega)}}\overline{\varphi'^{(2)}(\omega)}+\overline{\psi^{(2)}(\omega)} \ ,
\label{bccrack1conf}
\end{eqnarray}
\begin{eqnarray}
&&\frac{1}{\mu_1}[\kappa_1\varphi^{(1)}(\zeta)-\frac{\Lambda(\zeta)}{\overline{\Lambda'(\zeta)}}\overline{\varphi'^{(1)}(\zeta)}-\overline{\psi^{(1)}(\zeta)}]\nonumber\\&&=\frac{1}{\mu_2}[\kappa_2\varphi^{(2)}(\omega)-\frac{\Phi(\omega)}{\overline{\Phi'(\omega)}}\overline{\varphi'^{(2)}(\omega)}-\overline{\psi^{(2)}(\omega)}] \ .
\label{bccrack2conf}
\end{eqnarray}

\section{Method of Solution}
At this point we need to substitute the expansions (\ref{Laurentp1}),(\ref{Laurentp2}),(\ref{Laurent}) and an expansion similar to (\ref{Laurent}) for $\Lambda(\zeta)$ into the equations (\ref{bccrack1conf}) and (\ref{bccrack2conf}) and solve for the
coefficients $\varphi^{(i)}_k$ and $\psi^{(i)}_k$. To understand how to do this in principle we write the expanded equations (\ref{Laurentp1}) and (\ref{Laurentp2}) in an abstract form
\begin{equation}
\label{basic}
\sum_{k=-\infty}^{\infty}p_k\zeta^k=\sum_{m=-\infty}^{\infty}q_m\omega^m
\ ,
\end{equation}
where $p_k$ are linear combinations of the coefficients $\varphi^{(1)}_n$ and $\psi^{(1)}_n$ whereas
$q_m$ are linear combinations of $\varphi^{(2)}_n$ and $\psi^{(2)}_n$. As this equation stands we cannot use the orthogonality relation Eq. (\ref{ortho}). Therefore we expand moments of $\omega$ in terms of $\zeta$ in the form
\begin{equation}
\omega(\zeta)^m=\sum_{n=-\infty}^{\infty}a_{n,m}\zeta^n \ .
\label{momentexpansion}
\end{equation}
We now insert this expression in Eq. (\ref{basic}),
\begin{equation}
\label{doublef}
\sum_{k=-\infty}^{\infty}p_k\zeta^k=\sum_{m=-\infty}^{\infty}\sum_{n=-\infty}^{\infty}q_ma_{n,m}\zeta^n\
.
\end{equation}
and equate the coefficients of same powers to achieve a set of linear
equations for the coefficients of $\varphi^{(i)}$ and
$\psi^{(i)}$. The actual algebraic manipulations that are involved in reaching a {\em finite} set
of linear equations are presented in the appendix.
\begin{figure}
\epsfig{width=.35\textwidth,file=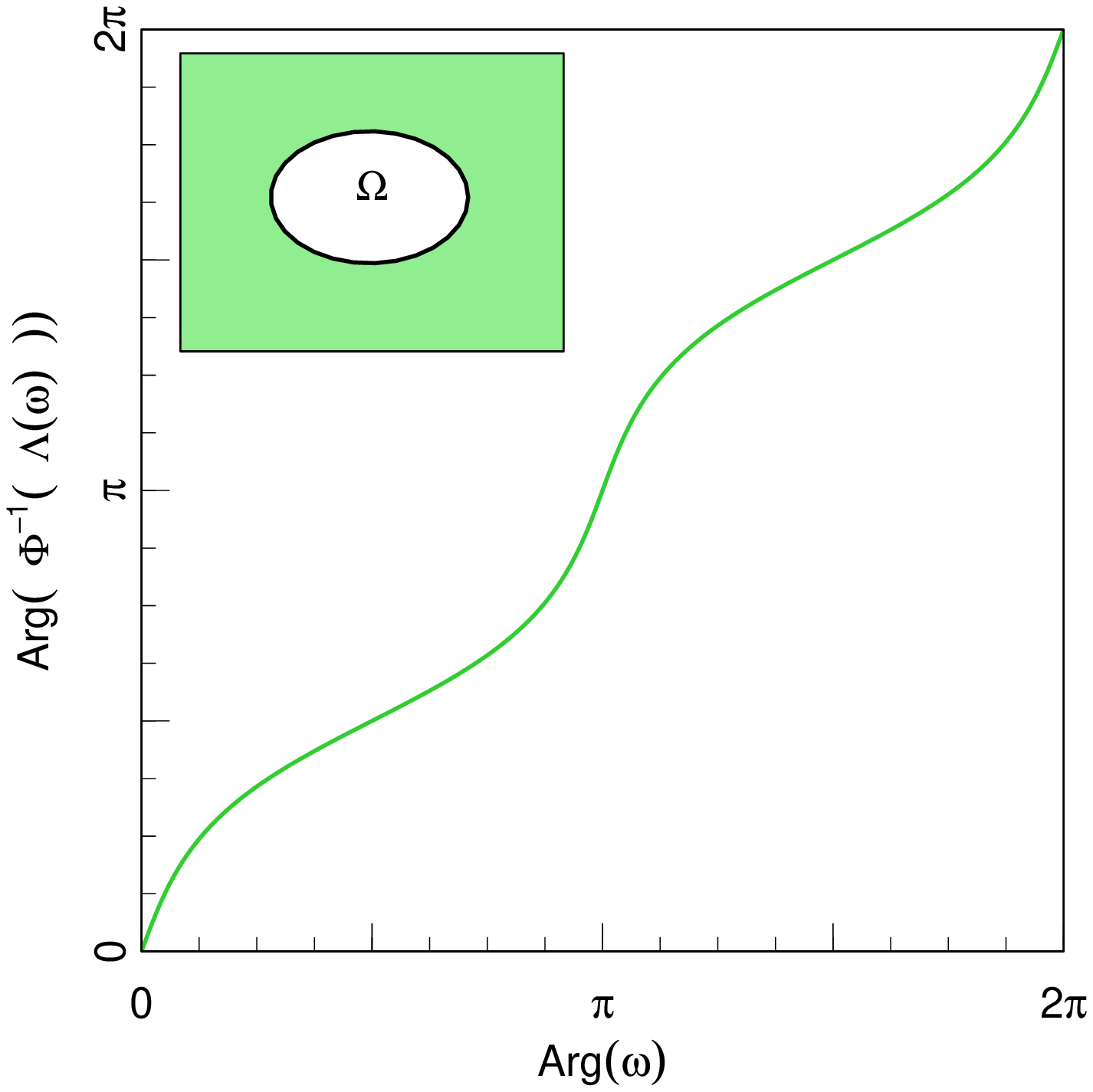,angle=-90}
\epsfig{width=.35\textwidth,file=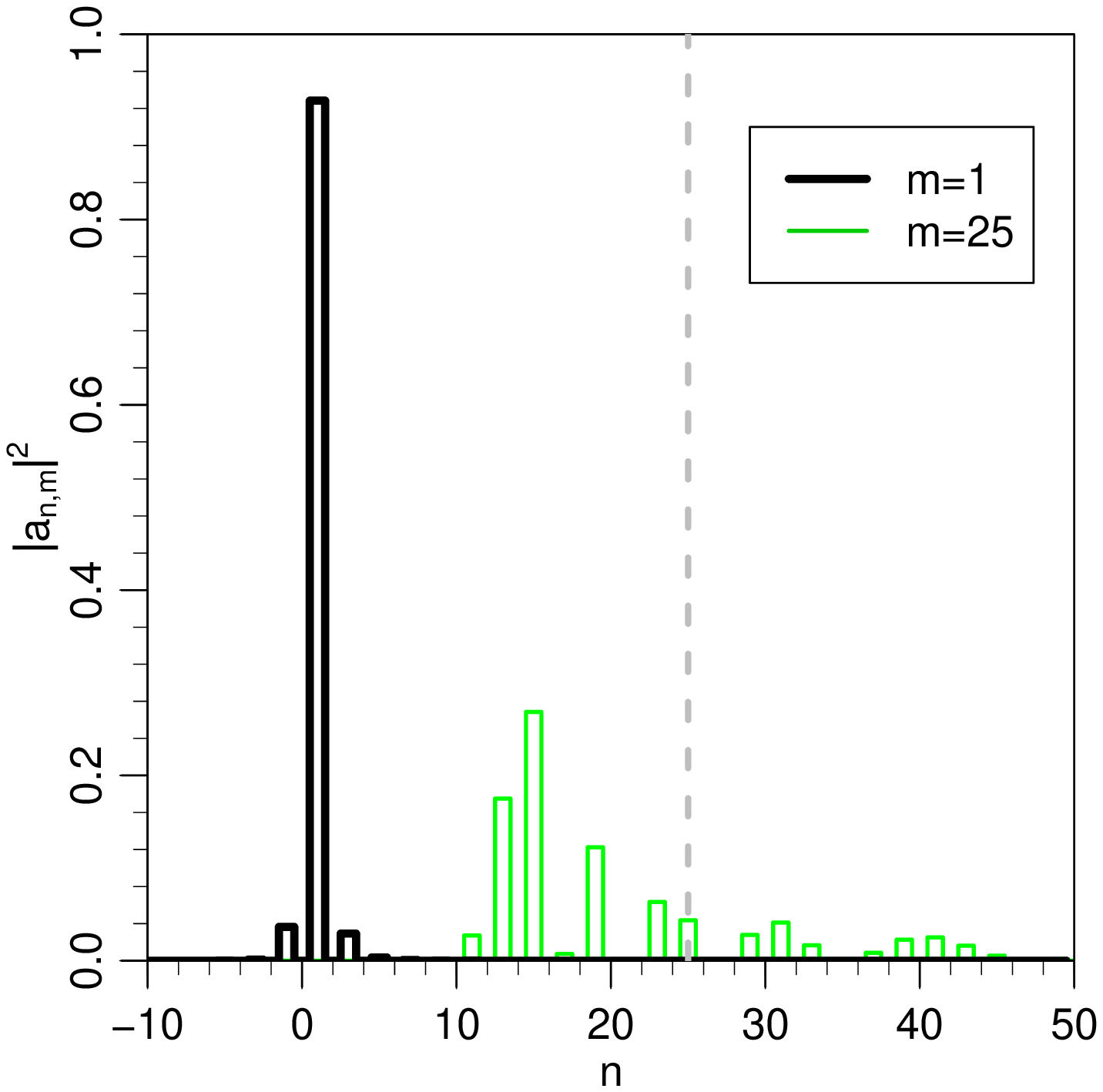,angle=-90}
\caption{In the upper panel, we show the relation between the mapping
  of the inner and outer domains of the ellipse drawn in the
  inset. The ellipse is the same as the one used in the Figs. \ref{figsxx} and \ref{figsyy}. Specifically, we have plotted the parametrization of the outer mapping as function of that of the inner, $\arg (\Phi^{-1}(\Lambda(\omega)))$. In the lower panel we show the power spectrum $|a_{n,m}|^2$ of the moments $m=1$ and $m=25$ of the function in the upper panel.}\label{t1t2}
\end{figure}

Needless to say, when we derive a finite set of equations we lose precision. To see this we note that to get the right number of equations for the number of unknowns (see Appendix) we need to truncate the summations on the LHS and the RHS of Eq. (\ref{doublef}) at the same finite N, i.e.
\begin{equation}
\label{doublefourier2}
\sum_{k=-N}^{N}p_k\zeta^k=\sum_{m=-N}^{N}\sum_{n=-N}^{N}q_ma_{n,m}\zeta^n\
.
\end{equation}
For a precisely circular inclusion this truncation introduces no loss of information.  For
this particular shape the
expansion Eq. (\ref{momentexpansion}) has only one term with
$n=m$, i.e. $a_{n,m}=\delta_{n,m}$. Obviously, when
the inclusion shape deviates from the circle, the representation of $\omega^m$ in Fourier space deviates from a delta function and it becomes more spread. An example of this phenomenon is
presented in  Fig.\ref{t1t2} for an inclusion in the form of an ellipse with aspect ratio of about 1.5.
The upper panel shows  the paramterization of the outer mapping as function of that of the inner, $\arg (\Phi^{-1}(\Lambda(\omega)))$.  In the lower panel we show the power spectrum $|a_{n,m}|^2$  of the moments $m=1$ and $m=25$ of the function in the upper panel. If we truncate the expansion at the dashed line $N=n=25$, we  lose high frequency information for the higher moments. This loss of informaion will lead to stress field calculations which are less accurate.
\begin{figure}
\epsfig{width=.35\textwidth,file=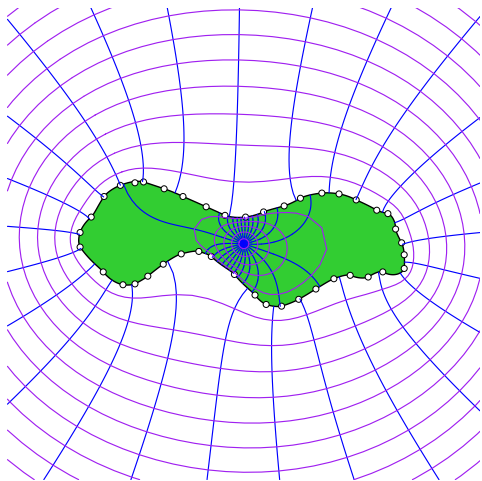}
\caption{Field lines of the two conformal mappings to the interior and exterior domains, respectively.}\label{figfield}
\end{figure}

To see the difficulty in a pictorial way we can consider
 the field lines of the conformal
mappings for an inclusion that is elongated in shape, see for example Fig.  \ref{figfield}. The external field lines concentrate at the convex parts of the inclusion, whereas the internal field lines concentrate on the concave parts.  It becomes increasingly difficult to match field lines since  they make a large
discontinues  jump when we go from the interior to the exterior
domain. Similarly, for the ellipse in Fig. \ref{t1t2}, when we increase
the aspect ratio of the ellipse, the slope in the steep parts of $\arg
(\Phi^{-1}(\Lambda(\omega)))$ become even larger, requiring higher order
frequencies  in our expansions. Eventually for
large aspect ratios, our method will break down.
\section{Obtaining the conformal maps}

In all the calculations we assumed that the conformal maps $\Phi(\omega)$ and $\Lambda(\zeta)$ are available. For arbitrary inclusion shapes this is far from obvious, and special methods are necessary to obtain these maps. An efficient method to obtain the conformal map from the exterior of the unit circle to the exterior of an arbitrary given shape had been discussed in great detail in \cite{06MPST}. In the present case we use a slightly different method namely the geodesic algorithm \cite{06MR}.  This method, like the former one, is based on the iterations of a generic conformal map $\gamma_\mathbf \xi$ defined by a set of parameters $\mathbf\xi$. We then construct the conformal map to an arbitrary shape by an appropriate choice of parameters $\mathbf\xi$. In the geodesic algorithm, we discretize the interface of the inclusion by a sequence of points $\{z_k \}_{k=0}^n$. The points appear sequentially in the positive direction of the interface.
\begin{figure}
\epsfig{width=.45\textwidth,file=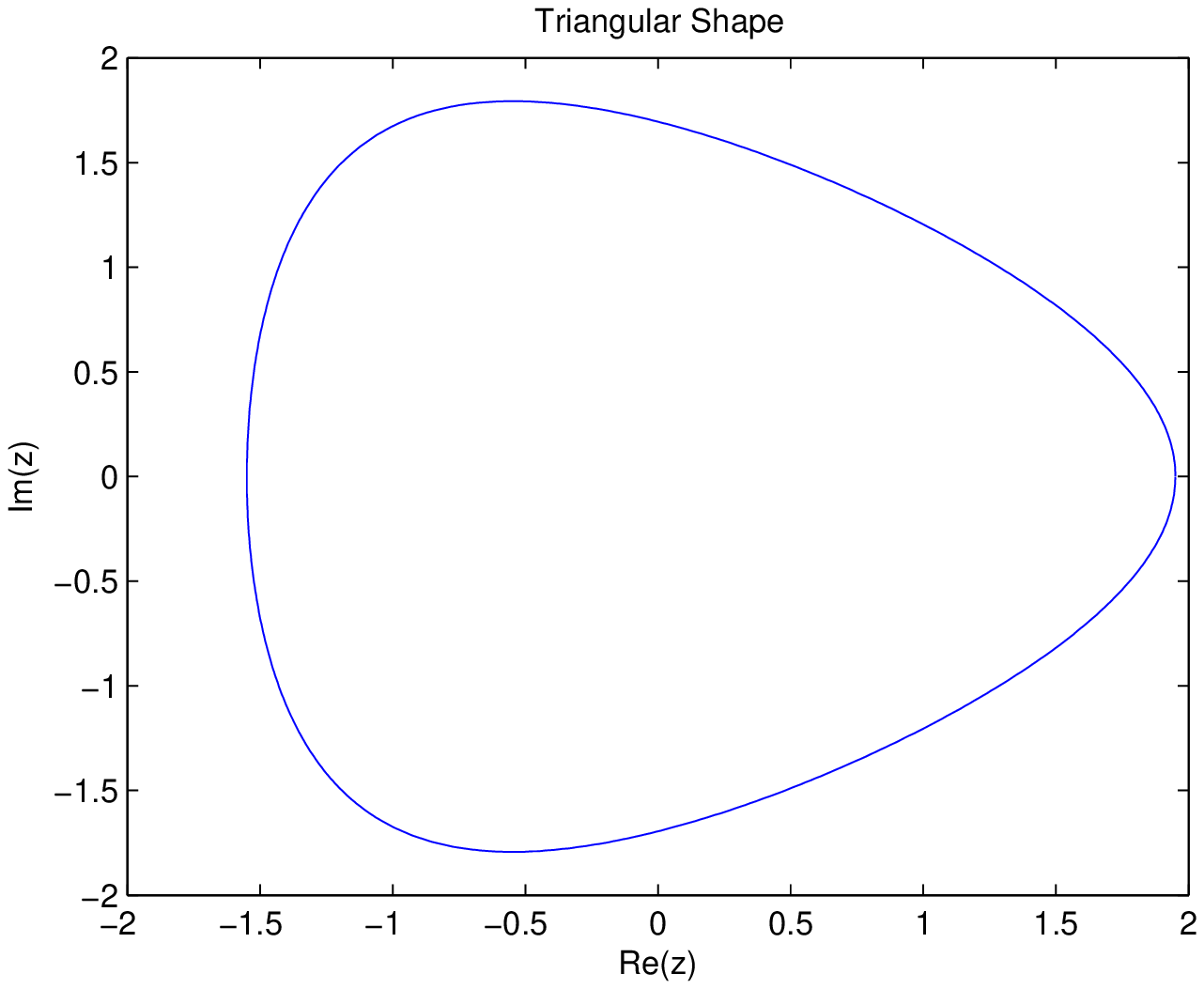}
\caption{}\label{triangle}
\end{figure}
\begin{figure}
\epsfig{width=.45\textwidth,file=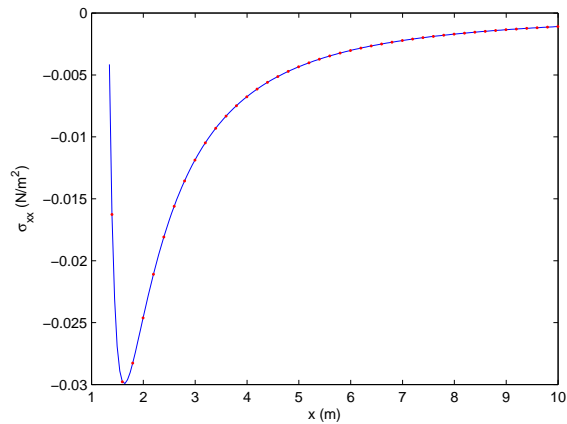}
\caption{$\sigma_{xx}$ evaluated along the positive real axis (ellipse)}\label{figsxx}
\end{figure}
We now briefly summarize how to construct the conformal map (see \cite{06MR}). First we construct iteratively the inverse map that brings the interior of the inclusion to the upper half plane and the interface to the real axis. The conformal map to the shape then follows directly from the inverse. The construction is done in three steps. In the first step, we move one point to infinity and another to the center of coordinates, e.g. $z_0$ and $z_1$, respectively. For that purpose we use the mapping,
$$
\gamma_1(z)=i\sqrt{\frac{z-z_1}{z-z_0}}
$$
In the next step we find a map that connects $z_2$ to the real axis by a semi circular arc. The inverse of this mapping, $\gamma_{\xi_2}$, brings $z_2$ to the real axis.
$$
\gamma_{\xi_2}(z)=\sqrt{\frac{z}{1-z/a}+b^2}
$$
where $\xi_2=\gamma_1(z_2)$ and $a=|\xi_2|^2/\Re \xi_2$ and $b=|\xi_2|^2/\Im \xi_2$.
Iteratively, we apply this mapping to all the points $z_3,\ldots z_{n}$ where in general
$$
\xi_k=\gamma_{\xi_{k-1}}\circ\ldots\circ\gamma_1(z_k).
$$
In the third and last step we unfold the remaining part of the interior to the whole upper half-plane by the map
$$
\gamma_{n+1}=-\left(  \frac{z}{1-z/\xi_{n+1}}\right)^2,
$$
with
$$
\xi_{n+1}=\gamma_{\xi_{n}}\circ\ldots\circ\gamma_1(z_0)
$$

The conformal map $\Psi$ from the upper half-plane to the interior domain and from the lower half-plane to the exterior domain is then given by
$$
\Psi=\gamma_1^{-1}\circ\gamma_{\xi_2}^{-1}\circ\cdots\circ \gamma_{\xi_{n}}^{-1}\circ\gamma_{n+1}^{-1} .
$$

\begin{figure}[h]
\epsfig{width=.45\textwidth,file=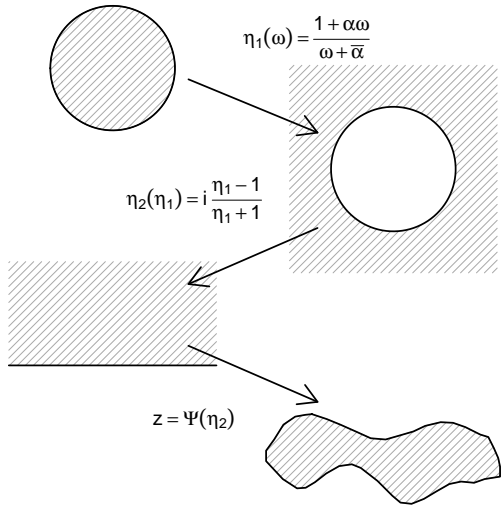}
\caption{Sketch of the construction of the conformal map from the exterior unit circle to the exterior of the inclusion. We choose $\alpha$ such that infinity is mapped to infinity. Similarly, for the interior map we choose $\alpha$ such that zero is mapped to zero.}\label{mapping}
\end{figure}

From the conformal map $\Psi$ we easily construct the map $\Phi$ from the unit circle to the inclusion. In Fig. \ref{mapping} we illustrate how this is done.

\section{Examples}
\begin{figure}
\epsfig{width=.45\textwidth,file=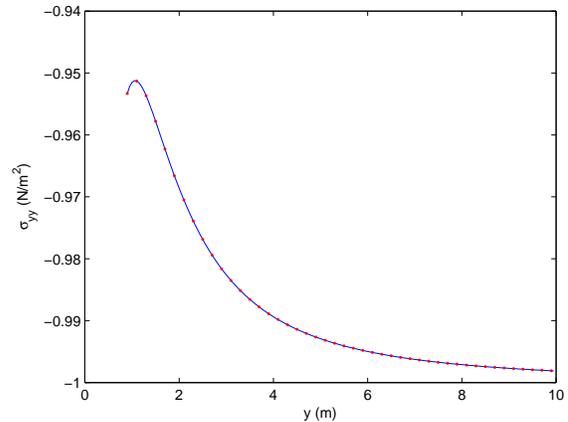}
\caption{$\sigma_{yy}$ evaluated along the positive imaginary axis (ellipse)}\label{figsyy}
\end{figure}

\begin{figure}
\epsfig{width=.45\textwidth,file=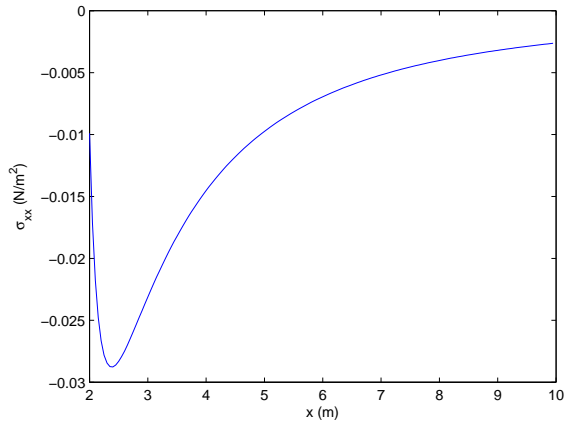}
\caption{$\sigma_{xx}$ evaluated along the positive real axis (triangle)}\label{trigxx}
\end{figure}

\begin{figure}
\epsfig{width=.45\textwidth,file=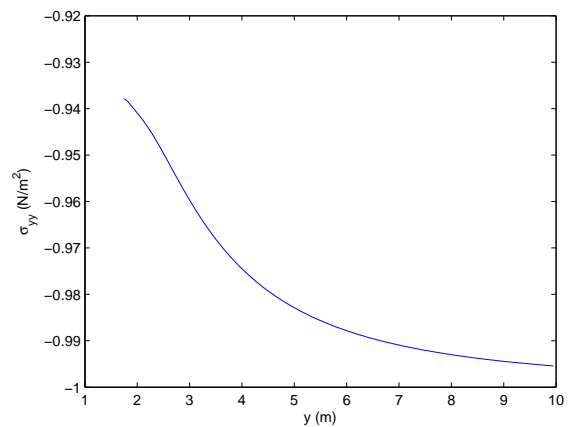}
\caption{$\sigma_{yy}$ evaluated along the positive imaginary axis (triangle)}\label{trigyy}
\end{figure}

In order to check the validity of our method we calculated the stress fields created by inhomogeneities with 2 different geometries: an ellipse with semi axes $0.9$ and $\sqrt{1+0.9^2}$ (aspect ratio of about 1.5) and a smoothed triangular curve $1.75z+\frac{0.2}{z^2}$ (see fig. \ref{triangle}).
In the case of the elliptical inhomogeneity we compared our method to the known analytical solution which was first obtained by Hardiman \cite{54Har} in 1954. In the example below, the
boundary conditions at infinity were set to
$\sigma_{\infty}=-1\frac{N}{m^2}$. The shear moduli used were
$\mu_1=1\frac{N}{m^2}$ for the inhomogeneity and
$\mu_2=1.2\frac{N}{m^2}$ for the matrix. The Poisson ratio was
taken to be $\nu=1$ for both inhomogeneity and matrix.
In figs. \ref{figsxx} and \ref{figsyy} we can see the components of the stress field calculated outside the ellipse.
The blue line is the stress calculated using Hardiman's solution and the red spots corresponds to the values obtained by our method.
Similarly, we have calculated the stress field outside the triangular-like inhomogeneity (figs \ref{trigxx} and \ref{trigyy}).
\section{Concluding Remarks}
In comparing our approach to other available algorithms, for example finite elements approximations
to the equations of linear elasticity, we should stress that our approach works equally well
for compressible and incompressible materials, There is no problem in taking the incompressible
limit as the Poisson ratio approaches 1/2. This is not the case for finite elements methods.
While the examples shown above worked out very well, indicating that the proposed
algorithm is both elegant and numerically feasible, unfortunately it
deteriorates very quickly when the shape of the inhomogeneity deviates
strongly from circular symmetry. The difficulty in matching the two
conformal maps is significant, as can be gleaned from from Figs. 2 and 3.
One could think that the problem could be overcome in principle by
increasing the numerical accuracy, but in practice, when the inhomogeneity
has horns, spikes or deep fjords, the difficulties becomes insurmountable.
Similar difficulties in another guise are however expected when any other
analytic or semi-analytic method is used, leaving very contorted
inhomogeneities as a remaining challenge for elasticity theory.

\acknowledgements
This work had been supported in part by the Israel Science Foundation, the German Israeli Foundation and the Minerva Foundation.
We thank Eran Bouchbinder, Felipe Barra, Anna Pomyalov and Charles Tresser for some very useful discussions.
\begin{widetext}
\appendix
\section{Explicit calculation}
\label{explicit}
Starting from (\ref{bccrack1conf}) and (\ref{bccrack2conf}) we
write the series expansions in a compact form:
\begin{equation}
\varphi^1(\zeta)=\sum_{k=0}^{\infty}\varphi^1_k \zeta^k, \quad
\psi^1(\zeta)=\sum_{k=1}^{\infty}\psi^1_k \zeta^k, \quad
\frac{\Lambda(\zeta)}{\overline{\Lambda'(\zeta)}}=\sum_{k=-\infty}^{\infty}b_k
\zeta^k \ ,
\end{equation}
\begin{equation}
\varphi^2(\omega)=\sum_{k=-\infty}^{1}\varphi^2_k \omega^k, \quad
\psi^2(\omega)=\sum_{k=-\infty}^{1}\psi^2_k \omega^k, \quad
\frac{\Phi(\omega)}{\overline{\Phi'(\omega)}}=\sum_{k=-\infty}^{\infty}c_k
\omega^k \ ,
\end{equation}
were we have eliminated the zero terms in $\psi^1$ because of the
gauge freedom. Substituting:
\begin{equation}
\sum_{k=0}^{\infty}\varphi^1_k
\zeta^k+(\sum_{n=-\infty}^{\infty}b_n\zeta^n)\sum_{k=0}^{\infty}k\varphi^{1*}_k
\zeta^{1-k}+\sum_{k=1}^{\infty}\psi^{1*}_k \zeta^{-k}= \nonumber
\end{equation}
\begin{equation}
=\sum_{k=-\infty}^{-1}\varphi^2_k
\omega^k+(\sum_{n=-\infty}^{\infty}c_n
\omega^n)\sum_{k=-\infty}^{-1}k\varphi^{2*}_k
\omega^{1-k}+\sum_{k=-\infty}^{0}\psi^{2*}_k\omega^{-k}+\varphi^2_1\omega+\frac{\Phi(\omega)}{\overline{\Phi'(\omega)}}\varphi^{2*}_1+\frac{\psi^{2*}_1}{\omega}\nonumber
\ .
\end{equation}
We can also expand:
\begin{equation}
\varphi^2_1\omega+\frac{\Phi(\omega)}{\overline{\Phi'(\omega)}}\varphi^{2*}_1+\frac{\psi^{2*}_1}{\omega}=\nonumber
\end{equation}
\begin{equation}
F_1\frac{\sigma_{\infty}}{4}\omega+\frac{\Phi(\omega)}{\overline{\Phi'(\omega)}}F_1\frac{\sigma_{\infty}}{4}+\frac{\sigma_{\infty}}{2}\frac{F_1}{\omega}=\sum_{n=-\infty}^{\infty}f_n\omega^n
\ .
\end{equation}
Substituting and changing the indices of summation: $m=1-k+n$ which leads to
$n=m+k-1$ we get
:
\begin{equation}
\sum_{m=0}^{\infty}\varphi^1_m
\zeta^m+\sum_{k=1}^{\infty}\sum_{m=-\infty}^{\infty}kb_{m+k-1}\varphi^{1*}_k
\zeta^{m}+\sum_{m=-\infty}^{-1}\psi^{1*}_{-m} \zeta^{m}=\nonumber
\end{equation}
\begin{equation}
=\sum_{m=-\infty}^{-1}\varphi^2_m
\omega^m+\sum_{k=-\infty}^{-1}\sum_{m=-\infty}^{\infty}
kc_{m+k-1}\varphi^{2*}_k
\omega^{m}+\sum_{m=0}^{\infty}\psi^{2*}_{-m}\omega^{m}+\sum_{m=-\infty}^{\infty}f_m\omega^m
\ .
\end{equation}
In order to find the relations between the coefficients we need
that both sides will be expressed in the same Fourier harmonics.
Therefore, we need to expand:
\begin{equation}
\omega(\zeta)=\chi(\Lambda(\zeta))\nonumber \ .
\end{equation}
In a Fourier series. The condition for this to be expanded in a
Fourier series is that $\omega(\zeta)$ is $L^2$ (i.e.
$\int_{-\pi}^{\pi}|\omega(e^{\imath\theta})|^2d\theta<\infty$ on
the segment $[-\pi,\pi]$). If that is the case, we can expand:
\begin{equation}
\omega(\zeta)=\sum_{k=-\infty}^{\infty}a_k\zeta^k\nonumber \ .
\end{equation}
Actually, it is found to be more convenient to expand powers of
$\zeta$ in a Fourier series:
\begin{equation}
\omega(\zeta)^m=\sum_{n=-\infty}^{\infty}a_{n,m}\zeta^n \ ,
\end{equation}
were $\zeta=e^{\imath\theta}$ and
$\int_{-\pi}^{\pi}|\omega(e^{\imath\theta})^m|^2d\theta<\infty$ on
the segment $[-\pi,\pi]$.
\begin{equation}
\sum_{n=0}^{\infty}\varphi^1_n
\zeta^n+\sum_{k=1}^{\infty}\sum_{n=-\infty}^{\infty}kb_{n+k-1}\varphi^{1*}_k
\zeta^{n}+\sum_{n=-\infty}^{-1}\psi^{1*}_{-n} \zeta^{n}=\nonumber
\end{equation}
\begin{equation}
=\sum_{n=-\infty}^{\infty}[\sum_{m=-\infty}^{-1}\varphi^2_m
a_{n,m}+\sum_{k=-\infty}^{-1}k(\sum_{m=-\infty}^{\infty}
c_{m+k-1}a_{n,m})\varphi^{2*}_k
+\sum_{m=0}^{\infty}\psi^{2*}_{-m}a_{n,m}+\sum_{m=-\infty}^{\infty}f_ma_{n,m}]\zeta^n
\ .
\end{equation}
Define:
\begin{equation}
B_{n,k}=\sum_{m=-\infty}^{\infty} c_{m+k-1}a_{n,m}
\end{equation}
and
\begin{equation}
C_n=\sum_{m=-\infty}^{\infty}f_ma_{n,m}
\end{equation}
Substituting:
\begin{equation}
\sum_{n=0}^{\infty}\varphi^1_n
\zeta^n+\sum_{k=1}^{\infty}\sum_{n=-\infty}^{\infty}kb_{n+k-1}\varphi^{1*}_k
\zeta^{n}+\sum_{n=-\infty}^{-1}\psi^{1*}_{-n} \zeta^{n}=\nonumber
\end{equation}
\begin{equation}
=\sum_{n=-\infty}^{\infty}[\sum_{m=-\infty}^{-1}\varphi^2_m
a_{n,m}+\sum_{k=-\infty}^{-1}kB_{n,k}\varphi^{2*}_k
+\sum_{m=0}^{\infty}\psi^{2*}_{-m}a_{n,m}+C_n]\zeta^n \ .
\end{equation}
Using the linear independence of the $\zeta^n$'s with respect to the Fourier integral, and 'cutting' the infinite series at some number N, we get a set of linear equations which is of the form:
\begin{equation}
\label{matform}
\hat{\mathbf{M}}_1\mathbf{v}=\mathbf{c}
\end{equation}
Where $\mathbf{v}$ is the vector of coefficients ($\varphi$'s and $\psi$'s), $\hat{\mathbf{M}}$ is a $(4N+2)\times(2N+1)$ matrix of constants and $\mathbf{c}$ is the $C_n$'s.
\newpage
Next, we substitute the expansions in the continuity equation for
the displacement:
\begin{equation}
\frac{1}{\mu_1}[\kappa_1\varphi^1(\zeta)-\frac{\Lambda(\zeta)}{\overline{\Lambda'(\zeta)}}\overline{\varphi'^1(\zeta)}-\overline{\psi^1(\zeta)}]=\frac{1}{\mu_2}[\kappa_2\varphi^2(\omega)-\frac{\Phi(\omega)}{\overline{\Phi'(\omega)}}\overline{\varphi'^2(\omega)}-\overline{\psi^2(\omega)}]\nonumber
\ . \label{bccrack}
\end{equation}
Substituting:
\begin{equation}
\frac{1}{\mu_1}[\kappa_1\sum_{k=0}^{\infty}\varphi^1_k
\zeta^k-(\sum_{n=-\infty}^{\infty}b_n\zeta^n)\sum_{k=1}^{\infty}k\varphi^{1*}_k
\zeta^{1-k}-\sum_{k=1}^{\infty}\psi^{1*}_k \zeta^{-k}]=\nonumber
\end{equation}
\begin{equation}
=\frac{1}{\mu_2}[\kappa_2\sum_{k=-\infty}^{-1}\varphi^2_k
\omega^k-(\sum_{n=-\infty}^{\infty}c_n
\omega^n)\sum_{k=-\infty}^{-1}k\varphi^{2*}_k
\omega^{1-k}-\sum_{k=-\infty}^{0}\psi^{2*}_k\omega^{-k}+\kappa_2\varphi^2_1\omega-\frac{\Phi(\omega)}{\overline{\Phi'(\omega)}}\varphi^{2*}_1-\frac{\psi^{2*}_1}{\omega}]
\ .
\end{equation}
We can also expand:
\begin{equation}
\kappa_2\varphi^2_1\omega-\frac{\Phi(\omega)}{\overline{\Phi'(\omega)}}\varphi^{2*}_1-\frac{\psi^{2*}_1}{\omega}=\nonumber
\end{equation}
\begin{equation}
\kappa_2F_1\frac{\sigma_{\infty}}{4}\omega-\frac{\Phi(\omega)}{\overline{\Phi'(\omega)}}F_1\frac{\sigma_{\infty}}{4}-\frac{\sigma_{\infty}}{2}\frac{F_1}{\omega}=\sum_{n=-\infty}^{\infty}g_n\omega^n
\ .
\end{equation}
Substituting and changing the indices of summation: $m=1-k+n$ which leads to
$n=m+k-1$:
\begin{equation}
\frac{1}{\mu_1}[\kappa_1\sum_{k=0}^{\infty}\varphi^1_k
\zeta^k-\sum_{k=1}^{\infty}\sum_{n=-\infty}^{\infty}kb_{m+k-1}\varphi^{1*}_k
\zeta^{m}-\sum_{k=1}^{\infty}\psi^{1*}_k \zeta^{-k}]=\nonumber
\end{equation}
\begin{equation}
=\frac{1}{\mu_2}[\kappa_2\sum_{m=-\infty}^{-1}\varphi^2_m
\omega^m-\sum_{k=-\infty}^{-1}\sum_{m=-\infty}^{\infty}
kc_{m+k-1}\varphi^{2*}_k
\omega^{m}-\sum_{m=0}^{\infty}\psi^{2*}_{-m}\omega^{m}+\sum_{m=-\infty}^{\infty}g_m\omega^m]
\ .
\end{equation}
Expanding $\omega(\zeta)$ as before and defining:
\begin{equation}
D_n=\sum_{m=-\infty}^{\infty}g_ma_{n,m} \ ,
\end{equation}
we get:
\begin{equation}
\frac{1}{\mu_1}[\kappa_1\sum_{n=0}^{\infty}\varphi^1_n
\zeta^n-\sum_{n=-\infty}^{\infty}\sum_{k=1}^{\infty}kb_{n+k-1}\varphi^{1*}_k
\zeta^{n}-\sum_{n=1}^{\infty}\psi^{1*}_n \zeta^{-n}]=\nonumber
\end{equation}
\begin{equation}
=\frac{1}{\mu_2}\sum_{n=-\infty}^{\infty}[\kappa_2\sum_{m=-\infty}^{-1}\varphi^2_m
a_{n,m}-\sum_{k=-\infty}^{-1}kB_{m,k}\varphi^{*2}_k
-\sum_{m=0}^{\infty}\psi^{2*}_{-m}a_{n,m}+D_n]\zeta^n \ .
\end{equation}
When 'cutting' the infinite series in the same way, we get again a matrix equation (with the same dimensions) of the form:
\begin{equation}
\label{matform2}
\hat{\mathbf{M}}_2\mathbf{v}=\mathbf{d}
\end{equation}
Combining equations (\ref{matform}) and (\ref{matform2}) we get a $(4N+2)\times(4N+1)$ matrix equation:
\begin{equation}
\label{matform2}
\hat{\mathbf{M}}\mathbf{v}=\mathbf{e}
\end{equation}
where
\begin{equation}
\hat{\mathbf{M}}=\hat{\mathbf{M}}_1\oplus\hat{\mathbf{M}}_2
\end{equation}
and
\begin{equation}
\mathbf{e}=\mathbf{c}\oplus\mathbf{d}
\end{equation}
\end{widetext}

\end{document}